\documentclass[sigconf]{acmart}

\usepackage{listings}
\usepackage{multirow}
\usepackage{graphicx}
\usepackage{hyperref}
\usepackage{fancybox}
\usepackage{enumitem}
\usepackage{svg}

\setcopyright{none}

\renewcommand\footnotetextcopyrightpermission[1]{} 
\ccsdesc[500]{Security and privacy~Software security engineering}
\ccsdesc[500]{Software and its engineering~Software maintenance tools}

\title{Sniffing for Codebase Secret Leaks with Known Production Secrets in Industry}
\begin{document}

\author{Zhen Yu Ding}

\author{Benjamin Khakshoor}

\author{Justin Paglierani}

\author{Mantej Rajpal}

\begin{abstract}
Leaked secrets, such as passwords and API keys, in 
codebases were responsible for numerous security 
breaches. Existing heuristic techniques, such as pattern 
matching, entropy analysis, and machine learning, exist to 
detect and alert developers of such leaks. Heuristics, 
however, naturally exhibit false positives, which require 
triaging and can lead to developer frustration. We propose 
to use \emph{known production secrets} as a source of ground truth 
for sniffing secret leaks in codebases. We develop 
techniques for using known secrets to sniff whole 
codebases and continuously sniff differential code revisions.
We uncover different performance and security needs when 
sniffing for known secrets in these two situations in an 
industrial environment.
\end{abstract}


\keywords{application security, continuous testing, experience paper, 
hard-coded secrets, leak detection, production secrets, 
secret management, secure coding, static analysis}

\maketitle

\section{Introduction}

Code repositories may accidentally leak secrets, such 
as private keys, API tokens, or passwords. Such leaks
facilitated numerous past security exploits~\cite{c1, c2, c3}.
Existing tools for detecting leaked secrets often use
heuristic techniques such as regex patterns~\cite{c4, c5, c6, c7, c8, c10},
entropy analysis~\cite{c6, c7, c9, c10}, and machine learning~\cite{c11, c12}.
Heuristic techniques, however, can be prone to false
positives, thus requiring manual triaging of analysis output.
Moreover, a blocking mechanism to prevent developers
from ever uploading secrets into codebases requires a very
low false-positive rate; otherwise, developers will become
frustrated and lose confidence~\cite{c13}.

We propose a method for detecting secret leaks in 
codebases that aims to minimize false positives. We use
\emph{known production secrets} as a source of ground truth for
detecting the presence of secrets. We extract known 
production secrets from our secret managers---tools that 
centrally manage storage of and access to secrets.
Although sniffing for known secrets cannot detect leaks of
secrets not stored in a secret manager, we expect very few 
false positives with our technique of sniffing with known 
secrets.

We apply our method of sniffing for known secrets to 
(1) sniff whole codebases and (2) continuously sniff 
differential code revisions. We discovered different 
performance and security requirements when using known 
secrets to sniff for leaks in these scenarios, which lead to 
different approaches to sniff for known secrets in these two 
scenarios.

\section{Approach}

We develop a continuous differential code revision 
sniffer to rapidly detect and respond to secret leaks. We 
complement our continuous sniffer with a whole codebase 
sniffer to detect secrets leaked prior to adopting the 
continuous sniffer. We additionally intend to periodically run
the whole codebase sniffer to detect leaks missed by the 
continuous sniffer, since a developer may leak a secret
into our codebase before adding the now-leaked secret into 
a secret manager.

\subsection{Whole Codebase Sniffer}

We sniff for secret leaks in whole codebases by pattern 
matching against raw production secrets. The Whole 
Codebase Sniffer pulls production secrets from secret 
managers, constructs regex patterns for each secret, and 
attempts to find matches for these instructions in our codebase.

When constructing a regex pattern for a secret, we 
consider the possibility that a hard-coded secret might be 
interrupted by whitespace or string concatenations. We thus 
match any whitespace and up to 5 non-whitespace 
characters between each character in the secret.

Since our codebase is gigabytes in size, our Whole
Codebase Sniffer needs to be reasonably performant on 
such large inputs.

\subsection{Continuous Differential Revision Sniffer}

\begin{figure}
	\includegraphics[width=\linewidth]{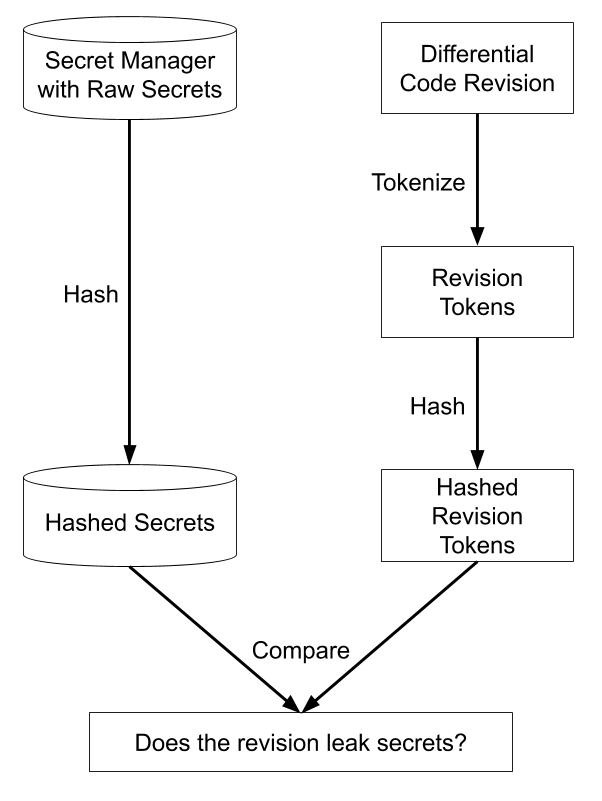}
	\caption{Continuous Differential Revision Sniffer}
	\label{fig:cdrs}
\end{figure}

Unlike the scenario of sniffing whole codebases, 
sniffing differential code revisions---such as git commits---does not require sifting 
through gigabytes of source code. However, our Continuous 
Differential Revision Sniffer needs to cope with a high
velocity of uploaded code revisions during times of busy
development.

We found the Whole Codebase Sniffer's step of pulling 
all production secrets from our secret managers per every 
invocation to be excessively slow for continuous sniffing. An 
alternative is to maintain a cache of raw production secrets
for our continuous sniffer. Caching our production secrets, however, 
increases our attack surface and requires strict 
access control to the system hosting the continuous sniffer.

To satisfy the performance needs of continuously 
sniffing a high velocity of code revisions while managing
risks to the confidentiality of our production secrets, we use 
a hashing-based approach to detect leaks of known secrets.
Figure~\ref{fig:cdrs} visualizes the process of sniffing for leaked secrets
in a differential code revision. The continuous sniffer
maintains a cache of hashed secrets derived from the raw
secrets in our secret managers. When a developer uploads 
a differential code revision for code review, the continuous
sniffer tokenizes the differential code revision. The sniffer 
then hashes the revision's tokens and compares the set of 
hashed tokens to the set of hashed production secrets. If 
the intersection of the two sets is nonzero, then the code
revision leaks secrets.

If the Continuous Differential Revision Sniffer detects
the leak of known secrets in a developer's code revision, 
the most secure next step is to block the developer from 
pushing the leaky code revision onto the remote codebase.
A developer-blocking solution, however, brings stricter
requirements for reliability and security of the sniffing tool 
and its supporting infrastructure. Failures in a 
developer-blocking secret sniffer, whether due to logic bugs 
or infrastructure failure, can---in the worst case---hinder every 
single software developer in the company from pushing 
code. Such failures can induce developers to lose trust and 
confidence in the company's security team. Such distrust can lead to 
an unwillingness to adopt developer-facing security tools in 
the future. Moreover, a developer-blocking sniffer would 
open a potential attack vector to brute force production 
secrets. Remedies to close this attack vector, such as rate 
limiting, are also subject to the same stringent reliability 
requirements of a developer-blocking sniffer. As an interim 
solution, the sniffer alerts security incident responders, rather than
block the developer, in the event of a detected secret leak.

\section{Conclusion and Future Work}

We propose using \emph{known production secrets} as a
source of ground truth for sniffing secret leaks with few false
positives. We intend to use known secrets to sniff both whole 
codebases and continuously sniff differential code revisions
in our industrial software engineering environment.

We intend to evaluate our sniffers on our internal 
codebases, and compare their performance to existing 
heuristics-based secret detection tools.

We observe that a substantial portion of our secrets 
reside in structured formats (e.g.: JSON, XML) alongside 
non-secret information. An example is a JSON credentials 
file containing a non-secret username and a secret
password. We may develop techniques to decompose such 
structured secrets in future work.

Since secret leaks are not exclusive to codebases, we 
are also considering extending our sniffers to inspect
emails, chat systems, and databases.

\bibliographystyle{ACM-Reference-Format}
\bibliography{references}

\end{document}